\documentclass[aps,prd,twocolumn,nofootinbib,superscriptaddress,amsmath,amssymb]
{revtex4-2}
\usepackage{silence} 
\usepackage{CJK}                     
\usepackage[dvips]{graphicx}         
\usepackage{mathptmx}                
\usepackage{bbm}
\usepackage{bm}
\usepackage{physics}
\usepackage{dcolumn}                 
\usepackage{multirow}                
\usepackage{xcolor}
\usepackage{hyperref}
\hypersetup{
breaklinks=true,colorlinks=true,
linkcolor=blue,citecolor=blue,filecolor=magenta,urlcolor=black}

\long\def\OFF#1{}
\long\def\hj#1{\color{red}#1\color{black}}

\def\be{\begin{equation}} \def\ee{\end{equation}}
\def\bal#1\eal{\begin{align}#1\end{align}}
\def\non{\nonumber}
 \def\pv{\bm{p}}
\def\ra{\rightarrow}
\def\eps{\varepsilon}
\def\phi{\varphi}
\def\tet{\theta}
\def\la{\lambda}
\def\al{\alpha}
\def\om{\omega}
\def\sig{\sigma}
\def\Om{\Omega}
\def\ms{\,M_\odot}
\def\mmax{M_\text{max}}
\def\mev{\,\text{MeV}}

\def\fm3{\,\text{fm}^{-3}}
\def\mfm{\,\text{MeV}\,\text{fm}^{-3}}

\def\khz{\,\text{kHz}}
\def\omb{\bar{w}}
\def\mc{\mathcal}
\def\ord{\mc{O}}

\def\sss#1{\vspace{-4mm}\subsubsection{#1}\vskip-3mm}

\begin{document}

\title{Quasiradial oscillations of rotating hybrid neutron stars}

\begin{CJK*}{UTF8}{gbsn}

\newcommand{\ccnu}
{Institute of Astrophysics, Central China Normal University,
Luoyu Road 152, Wuhan 430079, China}
\newcommand{\hue}
{School of Optoelectronic Information Engineering,
Institute of Astronomy and High Energy Physics,\\
Hubei University of Education,
Second Gaoxin Road 129, Wuhan 430205, China}
\newcommand{\cug}
{School of Mathematics and Physics, China University of Geosciences,
Lumo Road 388, Wuhan 430074, China}
\newcommand{\hust}
{School of Physics, Huazhong University of Science and Technology,
Luoyu Road 1037, Wuhan 430074, China}
\newcommand{\infn}
{INFN Sezione di Catania, Dipartimento di Fisica,
Universit\'a di Catania, Via Santa Sofia 64, 95123 Catania, Italy}

\author{Zi-Yue Zheng (郑子岳)}\affiliation{\ccnu}
\author{Ting-Ting Sun (孙婷婷)}\affiliation{\hue}
\author{\hbox{Huan Chen (陈欢)}}
\email{Email:huanchen@cug.edu.cn}\affiliation{\cug}
\author{\hbox{Xiao-Ping Zheng (郑小平)}}
\email{Email:zhxp@ccnu.edu.cn}\affiliation{\ccnu}\affiliation{\hust}
\author{\hbox{Jin-Biao Wei (魏金标)}}\affiliation{\cug}
\author{G. F. Burgio}\affiliation{\infn}
\author{H.-J. Schulze}\affiliation{\infn}

\begin{abstract}
We investigate fundamental quasiradial oscillations
in slow-rotation approximation
of pure and hybrid neutron stars,
employing equations of state
of nuclear matter from Brueckner-Hartree-Fock theory
or the relativistic mean field model,
and of quark matter from the Dyson-Schwinger quark model,
performing a Gibbs construction for the mixed phase in hybrid stars.
Characteristic differences between neutron-star
and hybrid-star fundamental quasiradial oscillation frequencies
during spin-down are pointed out.
\end{abstract}

\maketitle
\end{CJK*}

\section{Introduction}

Neutron stars (NSs) are among the densest objects known in the Universe
and serve as natural laboratories
for studying high-density nuclear matter (NM)
due to the extreme environment
shaped by the effects of the four fundamental interactions.
The interior of a NS can reach several times the nuclear saturation density
$\rho_0 \approx 0.16\fm3$.
At such high density,
the nucleons might undergo a phase transition to quark matter (QM),
and a hybrid star (HS) with a QM core is formed
\cite{Annala20}.

Theoretically,
the interior of a NS is characterized by the equation of state (EOS)
that links pressure and density.
Unfortunately,
due to the lack of precise computational methods
for nonperturbative strong interactions,
the high density in the cores of NSs poses a significant challenge
in developing a EOS for NSs
that accurately describes the behavior of dense matter under various conditions.
There are many theoretical models for the NS EOS
that satisfy the observational and experimental constraints;
for recent reviews,
see \cite{Burgio21,Chatziioannou25}.
For NM,
popular EOSs are based on
Brueckner-Hartree-Fock (BHF) theory
\cite{Li08b,Kohno13,Fukukawa15,Lu19,Wei21,Liu22},
relativistic mean field (RMF) models
\cite{Ring96,Hornick18,Shen20},
phenomenological models derived from energy-density functional theory
with generalized Skyrme effective forces
\cite{Potekhin13},
the self-consistent Green's functions approach
\cite{Carbone13},
the variational method
\cite{Akmal98},
and chiral effective field theory
\cite{Hebeler10,Coraggio14}.
For QM,
one currently still relies on more or less phenomenological models,
such as the MIT bag model
\cite{Chodos74},
the Nambu-Jona-Lasinio model
\cite{Schertler99,Buballa05,Klahn13,Klahn15,Li20,Liu23},
the perturbative quantum chromodynamics (QCD)
\cite{Kurkela09,Fraga14,Jimenez19,Somasundaram23,Finch25},
the quasiparticle model
\cite{Tian12,Zhao15,Li19,Li21},
and the quark-meson model
\cite{Zacchi15,Zacchi16}.
Moreover,
Dyson-Schwinger equations (DSEs)
provide a nonperturbative continuum field approach to QCD
that can simultaneously address
both confinement and dynamical chiral symmetry breaking
\cite{Roberts94,Alkofer01}.

After the groundbreaking direct observation of gravitational waves (GWs)
from a binary black hole (BH) merger \cite{Abbott16},
numerous additional GW signals have been recorded,
including those resulting from binary NS mergers
\cite{Abbott17a,Abbott20a}
and NS-BH mergers \cite{Abbott19a}.
These discoveries have paved the way for new investigations
into the internal structure of NSs,
significantly advancing our comprehension of dense NM \cite{Abbott18}
and the astrophysical phenomena
that occur under extreme conditions \cite{Abbott17b}.
For instance,
observations of GWs during the inspiral phase
have already imposed constraints on the EOS of NM at zero temperature,
revealing compatibility with estimates of radii and tidal deformabilities
\cite{Abbott18,Miller21,Pang21,Raaijmakers21,Rutherford24,Mauviard25,Miller26}.

Stellar oscillations provide a crucial window
into the internal structure of NSs
and can shed light on the complexities of high-density NM.
In particular,
the nonradial oscillations (NROs)
with spherical-harmonic decomposition orders of $l\geq2$
reveal details about the core of a NS through GWs \cite{Thorne67}.
Such emissions may occur
during a core-collapse supernova (CCSN) \cite{Radice19},
in the binary NS post-merger stage
\cite{Kokkotas01,Stergioulas11,Vretinaris20,Soultanis22},
or in an isolated perturbed NS \cite{Doneva13}.

For a nonrotating star,
the fluid modes of NROs include
$p$ (pressure) mode,
$f$ (fundamental) mode,
and $g$ (gravity) mode,
each characterized by distinct frequencies and damping times.
The $^2g_1$-mode eigenfrequencies are relatively small
and sensitive to new degrees of freedom inside the star,
such as deconfined QM
\cite{Weiwei20,Bai21,Jaikumar21,Constantinou21,Zhao22a,Zheng23},
hyperons
\cite{Dommes16,Tran23},
and superfluidity
\cite{Kantor14,Hang16,Dommes16,Rau18,Burgio21}.
In contrast,
the $p$-mode eigenfrequencies are too high to be observed
\cite{Kunjipurayil22},
exceeding the sensitivity range of next-generation GW detectors.
The fundamental $f$-mode eigenfrequencies of cold NSs
lie between those of $g$ and $p$ modes.
It is well established that the $f$-mode oscillation characteristics
exhibit universal relations with the stellar macroscopic features,
such as mean density or compactness
\cite{Andersson98,Kokkotas99,Pradhan22},
dimensionless tidal deformability
\cite{Chan14,Sotani21,Pradhan22,Zhao22b,Sotani23,Zheng25a,Zheng25b},
and dimensionless moment of inertia (MOI)
\cite{Lau09,Chirenti15,Zhao22b,Sotani23,Zheng25a,Zheng25b}.

Radial oscillation (RO),
being the simplest mode of stellar oscillation,
is routinely employed to assess stellar stability
against gravitational collapse.
In our previous work \cite{Sun21},
we found a clear difference between RO frequencies of pure NSs and HSs.
For nonrotating NSs,
although ROs do not directly produce GWs,
they might modulate the short gamma ray bursts \cite{Chirenti19}.

Rotation is a general characteristics of NSs,
influencing their structure.
The associated phenomena may serve as important features,
potentially enabling deduction of the properties related to the underlying EOS.
Due to rotational couplings,
the eigenfunctions of axisymmetric oscillation modes
will consist of a sum of various spherical harmonics \cite{Stergioulas03}.
Consequently,
quasiradial oscillations (QROs) could theoretically emit GWs,
which provides the possibility of observing QRO frequencies in GW signals.
\cite{Passamonti05,Passamonti07}
highlighted that signal amplifications are particularly significant,
i.e., resonance phenomena,
when the frequency of QROs approaches one of the axial $w$ modes of the star.
In CCSNe with rapidly rotating cores,
the deformations induced by rotation are known
to significantly enhance GW emission,
increasing the wave amplitude by a factor of 10 to 1000 \cite{Fryer11}.
The simulation of BH-forming CCSNe by \cite{Cerda13} suggests that
QROs could account for the turning point in frequency
occurring approximately 1.2~s after bounce,
as well as for the zero-frequency limit observed at BH formation.
\cite{Torres18} estimated that QROs are expected to emit GWs
with an amplitude of about $10\%$ of the quadrupole modes
for rapidly rotating cores.
Furthermore,
for the GWs from NS merger remnants,
\cite{Soultanis22} demonstrated that two side peaks of the dominant peak
originate from a nonlinear coupling between quadrupolar mode and QRO mode.

QROs in rotating NSs were initially studied
in the slow-rotation approximation \cite{Hartle75}.
Later,
\cite{Datta98} calculated the QRO frequencies
using the Chandrasekhar-Friedman formalism
\cite{Chandrasekhar72},
incorporating realistic EOSs.
\cite{Yoshida01} found that,
in the relativistic Cowling approximation,
apparent intersections between quasi-radial modes and other axisymmetric modes
can occur near the mass-shedding limit of rapidly-rotating stars.
Subsequently,
\cite{Font02} first investigated QROs for rapidly-rotating stars
in full general relativity
and compared with the results
obtained using the relativistic Cowling approximation.
\cite{Geroyannis14}
studied the impact of the polytropic index on the QRO frequencies
under the slow-rotation approximation.
\cite{Panda16} analyzed the effect of central density
on the QRO frequencies of pure NSs, HSs, and quark stars
in the presence of typical magnetic fields.

The spin-down of a rotating NS
can lead to a hadron-quark (HQ) phase transition
once a critical central density is reached
\cite{Glendenning97,Spyrou02,Ayvazyan13,Haensel16,Wei17}.
This phase transition may give rise to
a wealth of potentially observable signatures,
such as neutrino emission
\cite{Bhattacharyya06,Pagliara13},
gamma-ray bursts
\cite{Fryer98,Bombaci00,Bhattacharyya06,Mallick14},
GWs
\cite{Marranghello02,Lin11},
and electromagnetic radiation
\cite{Jaikumar07,Chen13}.
Notably,
previous studies have demonstrated that
such phase transitions could produce an observable signal
in the braking index of spin-down stars,
arising from a backbending behavior in the moment of inertia
\cite{Glendenning97,Spyrou02,Haensel16,Prasad22}.
However,
in our model \cite{Wei17},
the backbending phenomenon is not associated
with the phase transition via Gibbs construction,
but it occurs in supramassive stars prior to their collapse into BHs.
Detection of these signals may corroborate the existence of HSs
and illuminate the properties of high-density matter.

In this work,
we extend our investigation to explore the effects of rotation
on the ROs of NSs,
building upon our previous studies of nonrotating NSs \cite{Sun21}.
For NM,
we employ the BHF theory,
which is based on realistic two-nucleon interactions
and includes consistent microscopic three-nucleon forces
\cite{Grange89,Zuo02,Li08a,Li08b}.
The calculated saturation properties of NM using this approach
agree well with experimental data
\cite{Li08a,Li08b,Kohno13,Fukukawa15,Wei20,Burgio21}.
Furthermore,
the derived EOS simultaneously satisfies
the astrophysical constraints from the GW170817 event
and the mass-radius limits for NSs provided by NICER observations
\cite{Wei20,Burgio21,Wei21}.
To enhance the reliability of the results,
we conduct a comparative analysis using a modern RMF EOS \cite{Shen20},
which is characterized by a low symmetry energy slope.
For QM,
we adopt the Dyson-Schwinger model (DSM)
\cite{Chen11,Chen15},
which provides a continuum approach to quantum chromodynamics
that can simultaneously address both confinement
and dynamical chiral-symmetry breaking
\cite{Roberts94,Alkofer01}.
For the HQ phase transition,
we employ the Gibbs construction
\cite{Glendenning92,Chen11},
which determines a range of baryon densities
where hadron and quark phase coexist.
In this framework,
we investigate the QROs of NSs in the slow-rotation approximation.

This article is structured as follows.
In Sec.~\ref{s:eos},
we provide a brief overview of the formalism for the EOSs,
i.e., the BHF theory and the Shen RMF model for NM,
as well as the DSEs for QM.
In Sec.~\ref{s:osc},
we present the equilibrium structure equations
for both nonrotating and rotating stars,
along with the equations governing the QROs of NSs.
Numerical results are given in Sec.~\ref{s:res},
and we draw the conclusions in Sec.~\ref{s:end}.
We use natural units $G=c=\hbar=k_B=1$ throughout the article.

\section{Equation of state for neutron stars}
\label{s:eos}

\subsection{Nuclear matter}

The essential ingredient in the BHF many-body approach
is the in-medium reaction matrix $K$,
which is the solution of the Bethe-Goldstone equation
\be
 K(\rho,x_p;E) = V + \Re\sum_{1,2}V
 \frac{\ket{12} Q \bra{12}}{E-e_1-e_2} K(\rho,x_p;E) \:
\label{e:k}
\ee
and
\be
 U_1(\rho,x_p) = \Re\sum_{2<k_F^{(2)}}
 \expval{G(\rho,x_p;e_1+e_2)}{12}_a \:,
\label{eq:uk}
\ee
where $V$ is the nucleon-nucleon interaction potential,
$x_p \equiv \rho_p/\rho$ is the proton fraction,
and $\rho_p$ and $\rho$ are the proton and the total nucleon number densities,
respectively.
$E$ is the starting energy,
$Q$ is the Pauli operator, and
$e_i \equiv k_i^2\!/2m_i + U_i$ is the single-particle energy.
The multi-indices $1,2$ denote in general momentum, isospin, and spin.
The corresponding BHF energy density can then be expressed as
\be
 \eps = \sum_{1<k_F^{(1)}}
 \qty( \frac{k^2}{2m_1} + \frac12 U_1(k) ) \:,
\label{eq:f}
\ee
and the chemical potentials of the nucleons can be derived in a consistent way,
\be
 \mu_i = {\frac{\partial\eps}{\partial\rho_i}} \:.
\ee
We impose cold, neutrino-free, charge neutral, and catalyzed matter
consisting of neutrons, protons, and leptons ($e^-,\mu^-$)
in $\beta$ equilibrium due to weak interactions.
Finally, the pressure is given by
\be
 p(\eps) = \rho^2 \frac{\partial}{\partial\rho}
 \frac{\eps(\rho_i(\rho))}{\rho}
 = \sum_i \rho_i \mu_i - \eps \:.
\ee

The BHF approach requires only the potential $V$ as input.
In this study,
we employ the Argonne $V_{18}$ (V18) potential \cite{Wiringa95}
supplemented by compatible microscopic three-body forces
\cite{Grange89,Zuo02,Li08a,Li08b}.
This standard selection effectively reproduces
the saturation point and corresponding properties of symmetric NM.
For the energy density we adopt the convenient empirical parametrizations
provided in \cite{Lu19,Wei20}.

The nonrelativistic BHF theory predicts a relatively stiff EOS
characterized by a large sound speed,
which can even become superluminal at extremely high densities.
To address this causality violation,
we therefore also employ a RMF EOS,
which guarantees that the speed of sound remains below the speed of light.
Specifically,
we adopt the Shen 2020 EOS \cite{Shen20},
widely used in supernova simulations
\cite{Shen11,Sumiyoshi19}.
It is based on the TM1e parametrization \cite{Bao14},
which modifies the density dependence of the symmetry energy
by introducing additional $\omega$-$\rho$ coupling terms.
For this study,
we use the corresponding EOS table from the CompOSE database \cite{compose}.
Additionally,
the BHF approach only describes uniform NM in the core
and must be complemented by a crust EOS.
We adopt the same unified Shen 2020 EOS \cite{Shen20} for this purpose.
Since the core-crust transition density
has only minimal impact on the radius of NSs
with $M \geq 1.0\ms$ \cite{Burgio10,Baldo14,Fortin16},
we implement the simplest approach by ensuring continuity
of pressure and energy density at the transition interface.

\subsection{Quark matter}

For the description of deconfined QM,
we adopt the framework of DSEs
\cite{Chen11,Chen12},
which are the fundamental equations of motion
in continuum quantum field theory.
The basic quantity of the DSM is the quark propagator $S(p;\mu)$
at zero temperature and finite chemical potential $\mu$,
which can be expressed as
\be
 S(p;\mu)^{-1} = Z_2 \left[ i \bm{\gamma} \cdot \pv
 + i \gamma_4 \tilde{p}_4  + m_q \right] + \Sigma(p;\mu) \:,
\label{e:dse}
\ee
where $p=(\pv,p_4)$ is the four-momentum,
$\tilde{p}_4 \equiv p_4 + i\mu$,
and $m_q$ is the current mass of the quark $q=u,d,s$.
In this work,
we choose $m_u = m_d = 0$ and $m_s = 115\mev$.
The renormalized self-energy $\Sigma(p;\mu)$ is expressed as
\bal
 & \Sigma(p;\mu) =
\non\\
 & Z_1 g^2(\mu) \int\!\!
 \frac{d^4 q}{(2\pi)^4} D_{\rho\sig}(p-q;\mu)
 \frac{\la^a}{2} \gamma_\rho S(q;\mu)
 \frac{\la^a}{2} \Gamma_\sig(q,p;\mu) \:,
\label{e:dssigma}
\eal
where $Z_1$ and $Z_2$ are the quark-gluon-vertex and
quark-wavefunction renormalization constants, respectively.
$D_{\rho\sig}$ is the full gluon propagator,
$\Gamma_\sig$ is the full quark-gluon vertex,
$\la^a$ are the Gell-Mann matrices,
and $g(\mu)$ is the density-dependent coupling constant.
To obtain the quark propagator,
the so-called rainbow approximation and a
chemical-potential-modified Gaussian-type effective interaction
\cite{Chen11,Luo19} are adopted.

The number density, pressure, and energy density
for each quark flavor at zero temperature can be obtained as
\cite{Chen08,Klahn09}
\bal
 \rho_q(\mu_q) &= 6\int\!\! \frac{d^4p}{(2\pi)^4}
 \mathop{\text{tr}_D} \left[ -\gamma_4 S_q(p;\mu_q) \right] \:,
\label{e:dsrho}
\\
 p_q(\mu_q) &= p_q(\mu_{q,0})
 + \int_{\mu_{q,0}}^{\mu_q} d\mu \rho_q(\mu) \:,
\label{e:dsp}
\\
 \eps_q(\mu_q) &= -p_q(\mu_q) + \mu_q \rho_q(\mu_q) \:.
\label{e:dsed}
\eal
The total pressure and energy density are given
by summing contributions from all quark flavors
and those from the leptons.
The pressure of QM at zero density is determined by a
phenomenological bag constant \cite{Wei17},
\be
 B_\text{DS} = -\!\!\sum_{q=u,d,s} p_q(\mu_{q,0}) \:,
\label{e:BDS}
\ee
which is set to $90\mfm$ \cite{Chen12,Chen15,Wei17}.

In this work we adopt the Gibbs construction
\cite{Glendenning92,Chen11}
for the phase transition between hadron phase and quark phase.
The chemical and mechanical equilibrium in the mixed phase (MP)
are expressed as
\bal
 \mu_{B,H} &= \mu_{B,Q} \:,
\\
 \mu_{e,H} &= \mu_{e,Q} \:,
\\
 p_H(\mu_e,\mu_B) &= p_Q(\mu_e,\mu_B) = p_M(\mu_e,\mu_B) \:,
\eal
where the subscripts $H$, $Q$, and $M$
represent HM, QM, and the MP, respectively.
In the MP,
the local charge-neutrality condition is replaced by the global condition
\be
 \chi\rho_{C,Q} + (1-\chi)\rho_{C,H} = 0 \:,
\label{e:chi}
\ee
where $\chi$ is the volume fraction of QM in the MP.
Consequently,
the baryon number density $\rho_M$
and energy density $\eps_M$
of the MP can be determined as
\bal
 \rho_M &= \chi\rho_Q + (1-\chi)\rho_H \:,
 \\
 \eps_M &= \chi\eps_Q + (1-\chi)\eps_H \:.
\eal
Specifically, in the Gibbs construction
the pressure and energy density in HSs
are continuous functions of the baryon density,
at variance with the Maxwell phase transition \cite{Pereira18}.

\section{Neutron star structure}
\label{s:osc}

\subsection{Equilibrium configuration of spherical and static stars}

The Schwarzschild metric for a spherically-symmetric system is given by
\be
 ds^2 = - e^{\nu(r)} dt^2 + e^{\la(r)} dr^2
 + r^2( d\tet^2 + \sin^2\!\tet d\phi^2 ) \:,
\label{e:ds2}
\ee
where $e^{\nu(r)}$ and $e^{\la(r)}$ are metric functions.
The TOV equations \cite{Oppenheimer39,Tolman39}
for the hydrostatic equilibrium of stars in GR are given by
\bal
 p' &= - e^\la {q(m/r^2+4\pi rp)} \:,
\label{e:dpdr}
\\
 m' &= 4\pi r^2\eps \:,
\label{e:dmdr}
\\
 m_B' &= 4\pi r^2 e^{\la/2} \rho m_n \:,
\eal
where $p' \equiv dp/dr$ etc.,
$q \equiv p+\eps$, and
$m$ is the enclosed gravitational mass,
$m_B$ is the enclosed baryonic mass,
and $m_n$ is the neutron mass.
One can obtain radius $R$,
gravitational mass $M=m(R)$
and baryonic mass $M_B=m_B(R)$
of a NS for a given central pressure or density
by solving these equations with the boundary condition $p(R)=0$.
The corresponding metric functions are
\bal
 \nu' &= -\frac{2p'}{q} \:,
\label{e:nu}
\\
 e^\la &= \frac{1}{1-2m/r} \:.
\label{e:la}
\eal

\subsection{Structure of slowly rotating stars}

For a slowly rotating star with an angular velocity $\Om$,
the rotational effects can be treated perturbatively.
Following the framework developed in \cite{Hartle67,Hartle68,Hartle73},
the metric can be written as
\bal 
 ds^2 =& -e^{\nu(r)}[1+2h(r,\tet)] dt^2
 + e^{\la(r)} \left[ 1 + 2m^*(r,\tet) \right] dr^2
\non\\
 & + r^2 \left[ 1 + 2k(r,\tet) \right]
 \left\{ d\tet^2 + \sin^2\!\tet [d\phi-w(r,\tet)dt]^2 \right\}
\non\\
 & + \ord(\Om^4) \:.
\eal
Here $e^{\nu(r)}$ and $e^{\la(r)}$  
describe the nonrotating-star solution of the TOV equations,
as above.
The perturbative functions $h,m^*,k$, and $w$
can be expanded as \cite{Hartle73,Benhar05}
\bal
 h(r,\tet)   &= h_0(r) + h_2(r)P_2 + \ord(\Om^4) \:,
\\
 m^*(r,\tet) &= m_0(r) + m_2(r)P_2 + \ord(\Om^4) \:,
\\
 k(r,\tet) &= k_2(r)P_2 + \ord(\Om^4) \:,
\\
 w(r,\tet)   &= w_0(r) + w_1(r) - \frac{w_3(r)}{\sin\tet} \frac{dP_3}{d\tet}
 + \ord(\Om^5) \:,
\eal
where $P_l=P_l(\cos\tet)$ is the Legendre polynomial
and $v_2(r) \equiv k_2(r) + h_2(r)$ is introduced for simplicity.

\sss{First-order equations}

In this frame,
$\omb(r) \equiv \Om-w_0(r)$
represents the angular velocity of the fluid element relative
to the local inertial frame to order $\Om$.
It satisfies the following equation \cite{Hartle67}:
\OFF{
\be
 \frac{1}{r^4} \frac{d}{dr}\left(r^4 j \frac{d\omb}{dr}\right)
 + \frac{4}{r} \frac{dj}{dr} \omb = 0 \:,
\label{e:omb}
\ee
}
\be
 \left(r^4 j \omb' \right)' + 4 r^3 j' \omb = 0 \:,
\label{e:omb}
\ee
where $j(r) \equiv e^{-[\nu(r)+\la(r)]/2}$.
Starting with an arbitrary value for $\omb$
and $\omb'=0$ at $r=0$,
one can integrate this differential equation
and then match it at the surface $r=R$
to its analytic exterior solution
\be
 \omb(r) = \Om - \frac{2J}{R^3} \:,
\ee
where
\be
 J = \frac{R^4 \omb'(R)}{6} \:
\ee
is the angular momentum of the star.

\sss{Second-order equations}

For a rotating star,
the fluid element is displaced,
thus its energy density and pressure change.
In a reference frame that is momentarily moving with the fluid,
the pressure and energy density can be written as
\bal
    p(r,\tet) &= p(r) + q(r) \left[ p_0(r) + p_2(r) P_2 \right] \:,
\\
 \eps(r,\tet) &= \eps(r) + q(r) c_e^{-2}    
 \left[ p_0(r) + p_2(r) P_2 \right] \:,
\eal
where
$c_e^2=dp/d\eps$ is the squared equilibrium sound speed,
$p(r)$ and $\eps(r)$ denote the pressure and energy density
of the nonrotating background configuration, and
$p_0(r)$ and $p_2(r)$ are the monopole and quadrupole amplitudes
of the $\ord(\Om^2)$ changes of the pressure distribution.

Due to the $\ord(\Om^2)$ rotational corrections,
the shape of the star is distorted from a sphere into a spheroid.
The isodensity surface at radial coordinate $r$ in the nonrotating star
will in the rotating configuration
be displaced to
\be
 r + \delta r_0(r) + \delta r_2(r)P_2 + \ord(\Om^4)
\ee
with \cite{Hartle68}
\be
 \delta r_i(r)
 = \frac{2p_i}{\nu'} \:.
\label{e:dr}
\ee

\sss{Spherical deformation}

From the $l=0$ part of the Einstein field equation,
and knowing $\omb(r)$,
one can obtain the first-order differential equations
for $m_0(r)$ and $p_0(r)$ \cite{Hartle67,Hartle68},
\bal            
 m_0' &=
 4\pi r e^\la c_e^{-2} q p_0
 - \frac{e^\la}{r} \left( 1 - 8\pi r^2 \eps \right) m_0
\non\\&\quad
 + \frac{r^3 e^{-\nu}}{12} (\omb')^2
 + \frac{8\pi}{3} r^3 e^{\la-\nu}q\omb^2 \:,
\\
 p_0' &=
 - 4\pi r e^\la q p_0
 - \frac{e^\la}{r} \left(1+8\pi r^2 p\right) m_0
\non\\&\quad
 + \frac{r^3 e^{-\nu}}{12} (\omb')^2
 + \frac13 ({r^2 e^{-\nu} \omb^2} )' \:.
\eal
These two equations are integrated outward from the center of the star.
Assuming that the rotating configuration shares the same central density
as its nonrotating counterpart
implies the boundary conditions
$m_0(0) = 0$ and $p_0(0) = 0$
at the center.
At the surface, $p_0(R)=0$,
and the change of the gravitational mass of the star due to rotation
can be expressed as
\be   
 \delta M = R e^{-\la(R)} m_0(R) + \frac{J^2}{R^3} \:,
\label{e:dM}
\ee
whereas the change of the baryonic mass to $\ord(\Om^2)$ is
\cite{Hartle67,Hartle68}
\be   
 \delta M_B = m_n\!\! \int_0^R\!\! dr\; 4\pi r^2 e^{\la/2}
 \!\bigg[ \rho
 \bigg(\! m_0 + \frac{r^2 e^{-\nu} \omb^2}{3} \!\bigg)
 + \frac{d\rho}{dp} q p_0 \bigg] \:.
 \label{e:dMB}
\ee

The remaining monopole deformation $h_0$
is defined inside the star as \cite{Hartle68}
\be
 h_0 = h_{0,c} - p_0 + \frac{r^2 e^{-\nu} \omb^2}{3} \:,
\ee
where the constant $h_{0,c}$ is determined
by matching to the exterior solution at the surface,
\be   
 h_0(R) = -m_0(R) \:.
\ee

\sss{Quadrupole deformation}

The quadrupole moment depends on the $l=2$ perturbative functions
$h_2,m_2,v_2,p_2$.
According to Hartle's approach \cite{Hartle67},
the differential equations for $h_2$ and $v_2$ are
\bal
 h_2' =&\,
   \frac{r^4 j^2}{6} \bigg[ \frac{\nu'}{2} - \frac{e^\la}{r^2\nu'} \bigg]
 (\omb')^2
 - \frac{r^3 (j^2)'}{3} \bigg[ \frac{\nu'}{2} + \frac{e^\la}{r^2\nu'} \bigg]
 \omb^2
\non\\&\,
 + \bigg[ \frac{e^\la}{\nu'}
 \bigg( 8\pi q - \frac{4m}{r^3} \bigg) - \nu'  \bigg] h_2
 - \frac{4 e^\la}{r^2 \nu'} v_2 \:,
\label{e:dh2dr}
\\
 v_2' =&\,
 \bigg[ \frac{1}{r} + \frac{\nu'}{2} \bigg]
 \bigg[ \frac{r^4 j^2}{6} (\omb')^2
      - \frac{r^3 (j^2)'}{3} \omb^2 \bigg] - \nu' h_2 \:.
\label{e:dv2dr}
\eal
At the center, $r \ra 0$,
the solutions must be regular,
\be
 h_2(r) \ra Ar^2 \:,\quad
 v_2(r) \ra Br^4 \:,
\ee
where the constants are at $r=0$ related by
\be
 [3p+\eps] A + \frac{3}{2\pi} B = q [j\omb]^2 \:.
\ee
Outside the star,
Eqs.~(\ref{e:dh2dr},\ref{e:dv2dr})
can be solved analytically and the solution is
\bal
 h_2(r) &= \frac{J^2}{r^4} \left(\frac{r}{M} + 1\right)
 + K Q_2^2(r/M-1) \:,
\\
 v_2(r) &= -\frac{J^2}{r^4} + K \frac{2M e^{\la/2}}{r} Q_2^1(r/M-1)
\:,
\eal
where $Q_2^1$ and $Q_2^2$
are the associated Legendre functions of the second kind.
Then the constants $A$, $B$, and $K$ can be determined
by matching the boundary conditions at the surface.

The perturbative functions $m_2$ and $p_2$
can be obtained from the algebraic relations \cite{Hartle68}
\bal   
 m_2(r) &= \frac{r^4 j^2}{6} (\omb')^2
 - \frac{r^3(j^2)'}{3} \omb^2 - h_2 \:,
\\
 p_2(r) &= - \frac{r^2 e^{-\nu}}{3} \omb^2 - h_2 \:.
\eal

\sss{Third-order equations}

To order $\Om^3$,
the perturbation functions $w_1$ and $w_3$ will enter.
The corrections to angular momentum and moment of inertia
only depend on the function $w_1$,
which is determined by the second-order differential equation
\cite{Hartle73,Benhar05}
\be
 ( r^4 j w_1' )' + 4 r^3 j' w_1 = D_0 - D_2/5 \:,
\label{e:w1}
\ee
where
\bal   
 D_0 =& - r^4 j \big[ h_0 + m_0 \big]' \omb'
\non\\&
 - 4 r^3 j' \bigg[ 2m_0 + \frac{dq}{dp} p_0
                  + \frac{2}{3} r^2 e^{-\nu} \omb^2 \bigg] \omb \:,
\\
 D_2 =& - r^4 j \big[ 5 h_2 + m_2 - 4 v_2 \big]' \omb'
\non\\&
 - 4 r^3 j' \bigg[ 2m_2 + \frac{dq}{dp} p_2
                  - \frac{2}{3} r^2 e^{-\nu} \omb^2 \bigg] \omb \:
\eal
are given by the monopole and quadrupole amplitudes.
By introducing the variables \cite{Benhar05}
\be
 u_0 = j w_1 \:,\quad
 u_1 = r^4 j w_1' \:,
\ee
Eq.~(\ref{e:w1}) can be rewitten as the system
\bal
 u_0' & = - 4\pi r e^\la q u_0 + u_1/r^4 \:,
\\
 u_1' & = 16\pi r^4 e^\la q u_0 + D_0 - D_2/5 \:.
\eal
Inside the star,
the general solutions are linear combinations of particular solutions
and homogeneous solutions,
\be
 u_0 = u_0^P + C u_0^H \:,\quad
 u_1 = u_1^P + C u_1^H \:,
\ee
where $C$ is a constant.
The particular solutions can be obtained with the boundary conditions
$u_i^P(0) = 0$
at the center of the star.
Setting $D_0 = D_2 = 0$
yields the homogeneous equations,
with the asymptotic conditions as $r \ra 0$
\bal
 u_0^H(r) &\ra 1- \frac{2\pi}{5}q(0) r^2 \:,
\\
 u_1^H(r) &\ra \frac{16\pi}{5}q(0) r^5 \:.
\eal
Outside the star,
$u_0$ and $u_1$ can be solved analytically \cite{Benhar05},
\bal
 u_0(r) &= \frac{2 \delta J}{r^3}
 -\frac{12J^3}{5r^7} - \frac{4J^3}{5r^6 M} - \frac{33KJ}{40M^3}
\non\\&\quad
 + \frac{KJ}{40 r^4 M^3}
 \big[ 33 r^4 + 120 r^4 \la - 240 r^3 M - 288 r^3 M \la
\non\\&\quad
 + 336 r^2 M^2 + 256 r M^3 + 192 r M^3 \la - 96 M^4 \big] \:,
\\
 u_1(r) &= \frac{6J}{5} \bigg[ \frac{6J^2}{r^4} - 4k_2 \bigg] - 6\delta J \:,
\eal
where
$\delta J$ is the $\ord(\Om^3)$ correction to the angular momentum.
The constants $C$ and $\delta J$ can be obtained together,
matching the solutions of $u_0$ and $u_1$
at the surface of the star.
Finally,
the $\ord(\Om^2)$ contribution to the moment of inertia is
\be
 \delta I = \delta J / \Om \:.
\label{e:dI}
\ee

\subsection{Quasiradial oscillations}

The effects of slow rotation on the properties of a relativistic star
can be treated as perturbations
of the corresponding properties of a non-rotating star.
Therefore,
in the case of pulsations in a rotating star,
the squared oscillation frequency $\om^2$
can be expanded in powers of the angular velocity \cite{Hartle75}
\be
 \om^2 = \om_0^2 - \om_2^2 + \ord(\Om^4) \:.
\label{e:sig}
\ee

For the zeroth-order eigenfrequency
corresponding to the nonrotating case,
consider a spherically symmetric system with only radial motion.
In this scenario,
the metric Eq.~(\ref{e:ds2}) becomes time dependent.
Small perturbations are described by
$\xi \equiv \Delta r/r$,
where $\Delta r$ is the radial displacement,
and the corresponding Lagrangian perturbation of the pressure
$\eta \equiv \Delta p/p$.
The radial oscillation equations governing these perturbation functions
inside the star are \cite{Chanmugam77,Vath92}
\bal
 \xi' &= \bigg[ \frac{\nu'}{2} - \frac{3}{r} \bigg] \xi
 - \frac{1}{r\Gamma} \eta \:,
\label{e:dxidr}
\\
 \eta' &=
 r \frac{q}{p} \bigg[ \om_0^2 e^{\la-\nu} + \frac{2\nu'}{r}
 + \frac{(\nu')^2}{4} - 8\pi e^\la p \bigg] \xi
\non\\&\quad
 + \bigg[ \frac{\nu' \eps}{2p} - 4\pi r e^\la q \bigg] \eta \:,
\label{e:ddeltapdr}
\eal
where the adiabatic index is defined as
$\Gamma \equiv c_s^2 q/p$
with the squared adiabatic sound speed
$c_s^2 = \left(dp/d\eps\right)_s$,
$s$ being the entropy per baryon.
We also define the equilibrium adiabatic index
$\gamma \equiv c_e^2 q/p$
based on the EOS in $\beta$ equilibrium.
In this work,
$\Gamma=\gamma$,
since we adopt the EOSs for cold NS matter.

In order to determine the eigenfrequencies of oscillation,
one also needs boundary conditions,
which, in the NS center, are
\cite{Chanmugam77,Vath92}
\be
 3\Gamma\xi(0) + \eta(0) = 0 \:,
\label{e:r0}
\ee
while at the outer surface of the star,
the Lagrangian perturbation of the pressure should vanish
and Eq.~(\ref{e:ddeltapdr}) must be regular
\cite{Chanmugam77,Vath92},
\be
 \eta(R) = - \xi(R) \left[ 4 + e^{\la(R)}
 \left( \frac{M}{R} + \frac{\om_0^2 R^3}{M} \right) \right] \:.
\label{e:rR}
\ee
Eqs.~(\ref{e:dxidr}) to (\ref{e:rR})
are the Sturm-Liouville eigenvalue equations for $\om_0$.
Their solutions provide the discrete eigenfrequencies~$\om_0$.

The second-order shifts of the eigenfrequencies due to rotation
are computed by the relation \cite{Hartle72,Geroyannis14}
(Here we replace $U$ in Eq.~(4.8) of Ref.~\cite{Hartle72}
by $\xi = e^{\nu/2} U\!/r^3$)
\be
 \om^2_2 = - \frac
{ \int_0^R dr\; e^{(\la+\nu)/2} \mc{D}(r) \xi^2(r) }
{ \int_0^R dr\; e^{(3\la-\nu)/2} r^4 q \xi^2(r) } \:,
\label{e:sigma2}
\ee
where
\begin{widetext}
\bal   
 \mc{D}(r) &=
 \frac{ (r^3 e^{-\nu/2} \xi)' } { r^2 e^{-\nu/2} \xi}
 \bigg[ \mc{T}_0 + \frac23 r^4 e^{\la-\nu} \Gamma p \mc{T}_1 \omb^2 \bigg]
\non\\&\quad
 + e^{\la-\nu} \bigg[
   2 r^4 q h_0 \om_0^2
 + e^{\la+\nu} m_0 \mc{S}_0
 + r^2 e^{\la+\nu} q p_0 \mc{S}_1
 + \frac23 r^4  (\omb')^2 \mc{S}_2
 + \frac23 r^2 \omb^2 \mc{S}_3
 + 4 r^5 e^{-\la} \bigg( q + \frac{\Gamma}{3} p \bigg) \omb\omb'
 \bigg] \:,
\eal
and
\bal
 \mc{T}_0 &=
 r^2 e^\la \Gamma q m_0   
 + \frac{1}{2} r^2 (e^{-\la} - 1) \Gamma
   \bigg[ \frac{q}{\gamma p} - \frac{\eps}{p} \bigg] q p_0
 - \frac{1}{12} r^6 e^{-\nu} \Gamma q (\omb')^2
 - \frac{2}{3} r^5 e^{-\nu}
   \bigg[ \Gamma q + \frac{2(\Gamma p)^2}{q} \bigg] \omb\omb' \:,
\\
 \mc{T}_1 &= \frac{1}{2} (1-3e^{-\la}) \frac{q}{p}
 + \frac{1}{2} (1-5e^{-\la}) \Gamma \frac{p}{q}
 + \frac{1}{2} (1-e^{-\la}) \Gamma \bigg[ 1-\frac{1}{\gamma} \bigg]
 + 4\pi r^2 \Gamma \bigg[ 1 - \frac{1}{\gamma} + \frac{p}{q} \bigg] p
 - r e^{-\la} \Gamma' \frac{p} q \:,
\\
 \mc{S}_0 &=
 r^2 \Gamma q
 \Big[ (e^{-\la}-1)/2 + 4\pi r^2 (1+2e^{-\la}) p + (8\pi r^2 p)^2 \Big]
 - r^2 (q+\Gamma p)
 \Big[ 1+3e^{-\la} + 8\pi r^2 (2+e^{-\la}) p + (8\pi r^2 p)^2 \Big]
\non\\
 &\quad + r^3 e^{-\la} (1+8 \pi r^2 p) \Gamma' p
 - 2 r^4 e^{-\la-\nu} q \om_0^2 \:,
\\
 \mc{S}_1 &=
 \bigg[ \frac{\eps}{p} - \frac{q}{\gamma p} \bigg]
 \bigg[ r^2 e^{-\la-\nu} \om_0^2 + \frac14 (1-e^{\la}) (1+7e^{-\la}) \bigg]
 - 2\pi r^2 p \Big[ (1+e^{-\la}) (2+\Gamma) + 8\pi r^2 p (1+\Gamma) \Big]
 + 4\pi r^3 e^{-\la} \Gamma' p
\non\\&\quad
 - 2\pi r^2 q \Big[ (1-e^{-\la}) \Gamma
 + (1+e^{-\la}) (2-\Gamma)/\gamma
 + 8\pi r^2 p (1-\Gamma) (1+1/\gamma) \Big] \:,
\\
 \mc{S}_2 &=
 \pi r^4 (1 - \Gamma/2) p q
 + \pi r^4 \Gamma p^2
 + \frac{1}{16} r^2 (1-e^{-\la}) \Gamma q
 + \frac{1}{8} r^2 (1+7e^{-\la}) (q+\Gamma p)
 - \frac{1}{8} r^3  e^{-\la} \Gamma' p \:,
\\
 \mc{S}_3 &=
 - r^4 e^{-\nu} (q-\Gamma p) \om_0^2
 + \frac{r^2}{4} \Big[ 31e^{-\la} - 10 - e^\la + 2(e^{-\la}-1) \Gamma \Big]q
 + \frac{r^2}{4} \Big[ 6 - 11e^{-\la} + e^\la \Big] \Gamma p
\non\\&\quad
 + 2\pi r^4 (3+e^\la) (\Gamma-2) q p
 + 4\pi r^4 (1+e^\la) \Gamma p^2
 + (4\pi r^3 p)^2 e^\la [(\Gamma-1)q+\Gamma p]
 + r^3 e^{-\la} \Gamma' p\:.
\eal
\end{widetext}

To validate this complex code,
we performed calculations using the polytropic EOS
from Reference \cite{Hartle75}
and obtained consistent results.

\section{Numerical results}
\label{s:res}

\begin{figure}[t]
\vskip-5mm
\centerline{\includegraphics[width=0.5\textwidth]{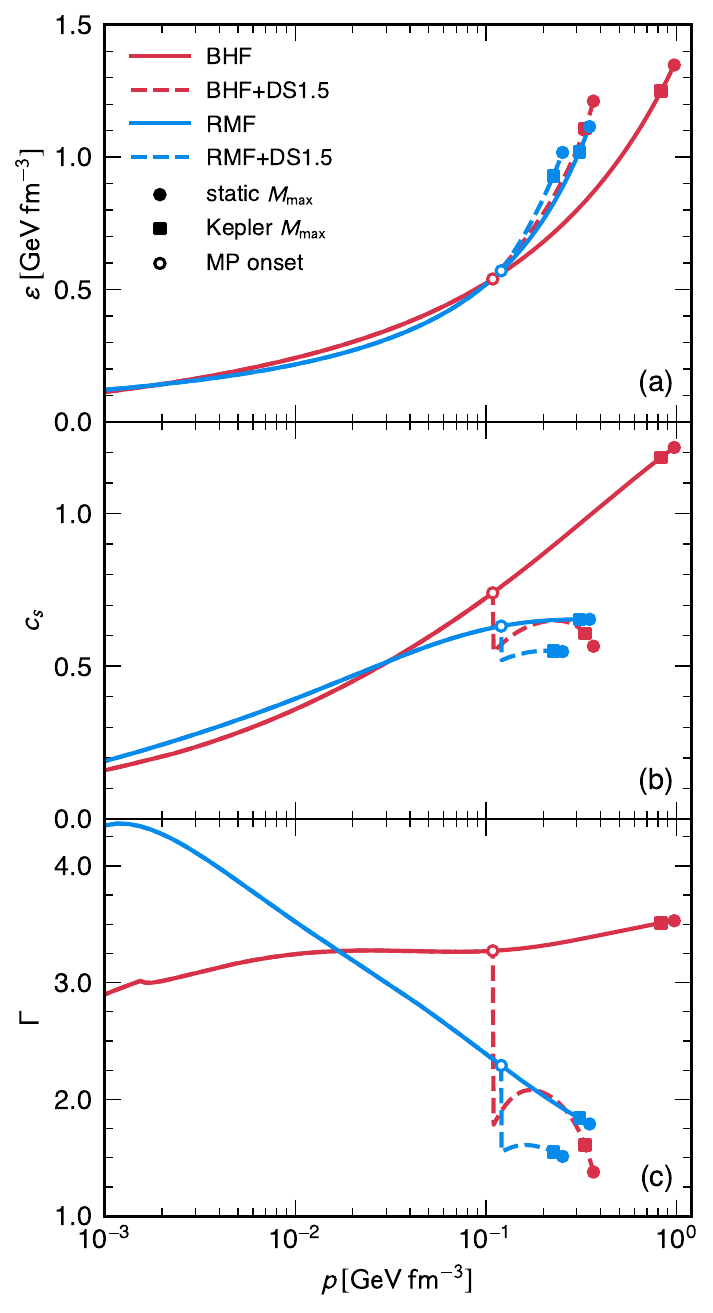}}
\vskip-5mm
\caption{
The energy density (a),
adiabatic sound speed (b), and
adiabatic index (c) of NSs
as functions of pressure with various EOSs.
Open markers indicate the onset of the MP.
The maximum-mass configurations
of static and Kepler-rotating stars
are indicated by full markers.
\hj{}
}
\label{f:eos}
\end{figure}

\begin{figure}[t]
\vskip-5mm
\centerline{\includegraphics[width=0.5\textwidth]{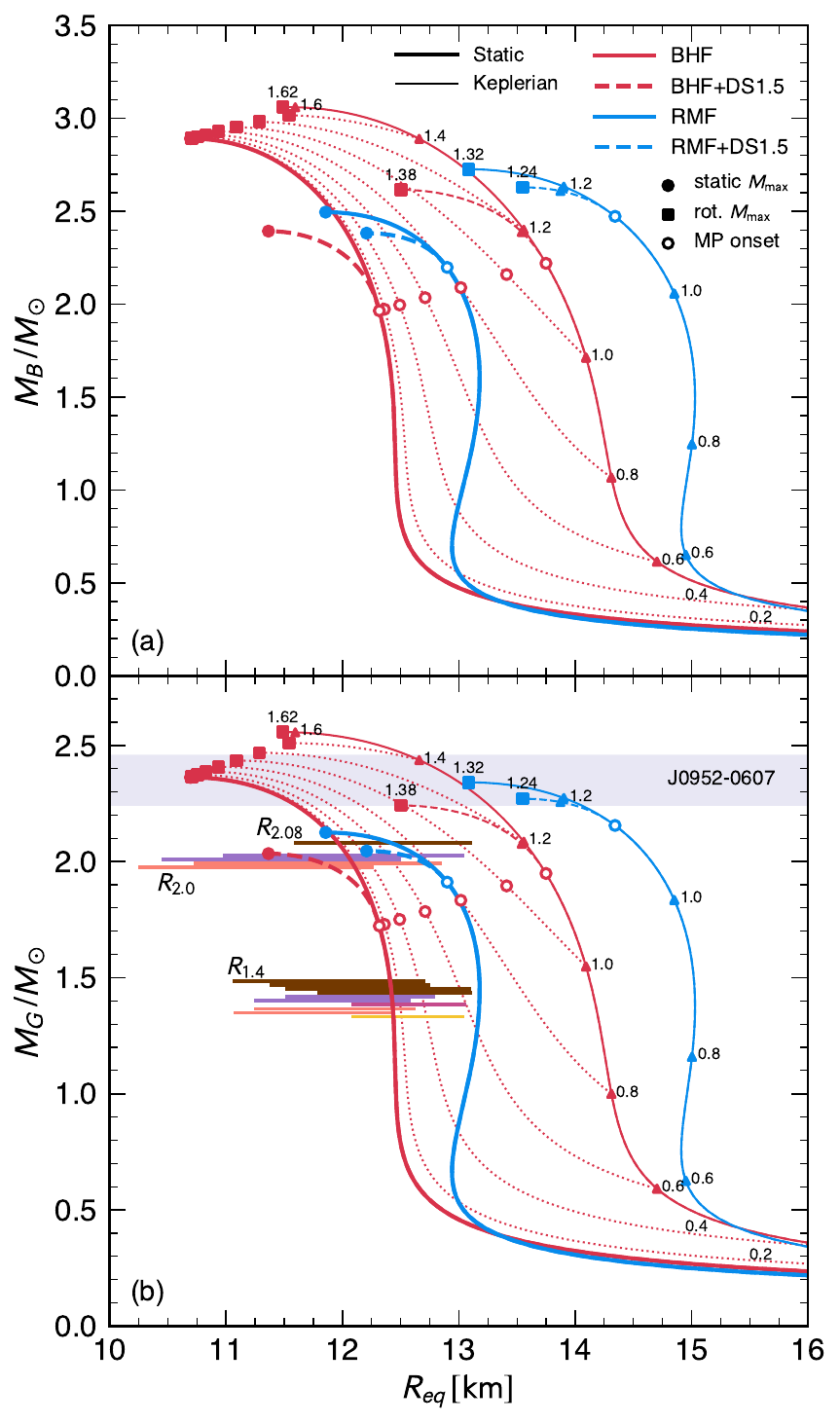}}
\vskip-5mm
\caption{
The baryonic (a) and gravitational NS mass (b)
versus equatorial radius for various EOSs.
Thick (thin) solid curves denote static (Keplerian) sequences.
Dotted curves are for fixed rotation frequency
($F=\Om/2\pi$ in kHz indicated near the curves).
The maximum-mass configurations are indicated by full markers.
Open dots indicate the bifurcation points of pure NSs and HSs.
The horizontal bars indicate the limits on
$R_{2.08}$, $R_{2.0}$, and $R_{1.4}$
obtained in the combined NICER+GW170817 data analyses
of \cite{Miller21,Pang21,Raaijmakers21}
(brown bars)
and the recent
\cite{Rutherford24,Dittmann24,Mauviard25,Miller26}
(other colored bars).
The mass range of the heaviest currently known NS
J0952-0607 \cite{Romani25} is also shown.
\hj{}
}
\label{f:m-r}
\end{figure}

\squeezetable
\begin{table}[t]
\def\myr#1{\multicolumn{1}{r}{$#1$}}
\def\myc#1{\multicolumn{1}{c}{$#1$}}
\renewcommand{\arraystretch}{1.4}
\setlength{\tabcolsep}{-2pt}
\caption{
Observational constraints on NS masses and radii.}
\begin{ruledtabular}
\begin{tabular}{lddr}
 Source      & \myr{M/\ms}   & \myc{R\ \text{[km]}} & Refs. \\
\hline\\[-4mm]
 J1614-2230  & 1.908\pm0.016 &                     & \cite{Arzoumanian18} \\
 J0348+0432  & 2.01\pm0.04   &                     & \cite{Antoniadis13}  \\
 J0348+0432  & 1.806\pm0.037 &                     & \cite{Saffer25}      \\
 J0740+6620  & 2.08\pm0.07   &                     & \cite{Fonseca21}     \\
 J0952-0607  & 2.35\pm0.11   &                     & \cite{Romani25}      \\
\hline
\multirow{5}{*}{J0030+0451}
 & 1.44^{+0.15}_{-0.14}    & 13.02^{+1.24}_{-1.06} & \cite{Miller19,Riley19} \\
 & 1.36^{+0.15}_{-0.16}    & 12.71^{+1.14}_{-1.19} & \cite{Riley21,Miller21} \\
 & 1.40^{+0.13}_{-0.12}    & 11.71^{+0.88}_{-0.83} & \cite{Vinciguerra24}    \\
 & 1.70^{+0.18}_{-0.19}    & 14.44^{+0.88}_{-1.05} & \cite{Vinciguerra24}    \\
 & 1.43^{+0.20}_{-0.17}    & 12.68^{+1.31}_{-1.04} & \cite{Kini26}           \\
\hline
\multirow{4}{*}{J0740+6620}
 & 2.08\pm0.07             & 13.7^{+2.6}_{-1.5}    & \cite{Riley21}      \\
 & 2.072^{+0.067}_{-0.066} & 12.39^{+1.30}_{-0.98} & \cite{Miller21}     \\
 & 2.073\pm0.069           & 12.49^{+1.28}_{-0.88} & \cite{Salmi24}      \\
 & 2.08\pm0.07             & 12.92^{+2.09}_{-1.13} & \cite{Dittmann24}   \\
\hline
\multirow{2}{*}{J0437-4715}
 & 1.418\pm0.037           & 11.36^{+0.95}_{-0.63} & \cite{Choudhury24}  \\
 & 1.418\pm0.044           & 13.7\pm1.8            & \cite{Miller26}     \\
\hline
\multirow{1}{*}{J0614-3329}
 & 1.44^{+0.06}_{-0.07}    & 10.29^{+1.01}_{-0.86} & \cite{Mauviard25}   \\
\hline
\multirow{1}{*}{J1231-1411}
 & 1.04^{+0.05}_{-0.03}    & 12.6\pm0.3 \text{\ or\ }
                             13.5^{+0.3}_{-0.5}    & \cite{Salmi24b}     \\
\hline
\multirow{7}{*}
{$\displaystyle\text{GW170817}\atop\displaystyle\text{+NICER}$}
 & 2.08                    & 12.35^{+0.75}_{-0.75} & \cite{Miller21}     \\
 & 2.0                     & 12.33^{+0.70}_{-1.34} \text{\ or\ }
                             11.55^{+0.94}_{-1.09} & \cite{Rutherford24} \\
 & 2.0                     & 11.99^{+0.85}_{-1.25} \text{\ or\ }
                             11.20^{+1.05}_{-0.94} & \cite{Mauviard25}   \\
 & 1.4                     & 12.28^{+0.50}_{-0.76} \text{\ or\ }
                             12.01^{+0.56}_{-0.75} & \cite{Rutherford24} \\
 & 1.4                     & 12.05^{+0.56}_{-0.79} \text{\ or\ }
                             11.71^{+0.71}_{-0.63} & \cite{Mauviard25}   \\
 & 1.4                     & 12.57^{+0.49}_{-0.48} & \cite{Dittmann24}   \\
 & 1.4                     & 12.56^{+0.47}_{-0.47} & \cite{Miller26}     \\
\end{tabular}
\end{ruledtabular}
\label{t:obs}
\end{table}

\subsection{Equation of state}

As discussed above,
we adopt the BHF EOS with the V18 potential
and the Shen 2020 RMF EOS for NM.
For the QM DSM EOS in HSs,
the free parameter $\al$
quantifies the strength of the in-medium modification
with a Gaussian-type effective interaction.
Here we select $\al = 1.5$
and combine it with the BHF and RMF EOSs,
labeled as BHF/RMF+DS$\al$, respectively,
to satisfy the constraint $\mmax > 2\ms$.

Fig.~\ref{f:eos} displays
the energy density (a),
adiabatic sound speed (b),
and adiabatic index (c)
as functions of pressure.
The maximum-mass configurations for static stars
and the onset of the MP
are marked by solid and open dots, respectively.
In panel (a) one can see that the EOS of the MP (dashed curves)
is generally softer than that of pure NM (solid curves),
while pure QM appears at densities too high
to be reached in HSs within our approach.

For pure NM,
the differing curvatures of the EOS $\eps(p)$
lead to notable variations in the adiabatic speed of sound,
as shown in panel (b).
The sound speed increases markedly in the BHF EOS,
while it stabilizes around $c_s \approx 0.65$ in the RMF model.
Such differences are reflected in the adiabatic index presented in panel (c),
which remains elevated for the BHF EOS,
but decreases with pressure for the RMF model.
In the MP,
the softening of the EOS results in a lower adiabatic speed of sound
and prevents the violation of causality encountered in the BHF model.
Similarly, the adiabatic index drops to a lower value.

\subsection{Stellar structure}

The mass-radius relations of pure NSs and HSs
are shown in Fig.~\ref{f:m-r}
with the same legend as in Fig.~\ref{f:eos}.
The lower panels show the gravitational mass as function of equatorial radius,
while the upper panels show the corresponding baryonic mass,
useful for studying the spin-down evolution of NSs later.
The dashed curve segments indicate the HS branches.
All EOSs considered in this article are compatible
with the two-solar-mass constraint
\cite{Arzoumanian18,Antoniadis13,Fonseca21}.
However, the low maximum masses of the RMF model
and the hybrid models might conflict with
the recent improved measurement $M=2.35\pm0.11\ms$
for PSR J0952-0607 \cite{Romani25},
while $\mmax \approx 2.36\ms$ of the BHF EOS is fully compatible.

The EOSs are also compatible with
the mass-radius results of the NICER mission for the pulsars
J0030+0451,
J0740+6620,
J0437-4715,
J0614-3329,
and PSR J1231-1411,
as summarized in Table~\ref{t:obs}.
The more constraining tightened limits on
$R_{2.08}$ \cite{Miller21},
$R_{2.0}$ \cite{Rutherford24,Mauviard25}, and
$R_{1.4}$
\cite{Pang21,Raaijmakers21,Rutherford24,Dittmann24,Mauviard25,Miller26},
from combined analyses of those pulsars together with GW170817,
are shown as horizontal bars in Fig.~\ref{f:m-r}(b).
The BHF EOS aligns well with these constraints
\cite{Sun21,Burgio21,Wei20},
while the RMF EOS complies only marginally with the $R_{1.4}$ range
in particular.

In this perturbative calculation,
we first choose a central baryon density
and integrate the TOV equations for a given EOS,
which yields the structure of the static, spherically symmetric background.
Keeping the central baryon density fixed,
we then assign a rigid angular frequency $\Om$ to the star
and compute the first-, second-, and third-order corrections in $\Om$
to the quantities of interest.

However,
the slow-rotation approximation breaks down
as the star approaches the Keplerian limit,
in particular,
the Keplerian angular velocity itself
cannot be calculated perturbatively.
Therefore we adopt here the approximation
\cite{Haensel09,Haensel16,Koliogiannis20}
\be
 \Om_K
 \approx 0.65 \sqrt{\frac{M_G}{R^3}}
\approx 7.48\times10^3\:\text{rad s}^{-1}
 \left[ \frac{M_G}{\ms} \right]^{1/2}
 \left[ \frac{R}{10\:\text{km}} \right]^{-3/2}  
\label{e:f_K}
\ee
with $M_G$ and $R$ obtained from the static TOV equations
as mentioned above,
to compute the maximum rotational $\ord({\Om^2})$ corrections to the
gravitational mass $\delta M$, Eq.~(\ref{e:dM}),
baryonic mass, Eq.~(\ref{e:dMB}),
equatorial radius
$\delta R_\text{eq}
= \delta r_0(R) + \delta r_2(R) P_2(0)
= \delta r_0(R) - \delta r_2(R)/2$,
Eq.~(\ref{e:dr}),
and moment of inertia $\delta I$, Eq.~(\ref{e:dI}).
We have verified that Eq.~(\ref{e:f_K})
agrees very well with result obtained by the RNS code
\cite{Stergioulas95,RNS_code,Wei17,Figura21}
in full GR.
Note again that at $\Om=\Om_K$ the $\ord({\Om^2})$ results are not converged
and very unreliable.
However,
the masses and radii for smaller frequencies
$\Om\ll\Om_K$,
where the slow-rotation approximation is valid,
can be easily obtained by a rescaling,
$\delta M(\Om) = (\Om/\Om_K)^2 \delta M(\Om_K)$ etc.

The corresponding $M(R)$ results are shown as thin solid (Keplerian)
and dotted (fixed $F=\Om/2\pi$) curves in Fig.~\ref{f:m-r}.
The maximum (approximate) Keplerian frequencies are about 1.6~kHz,
in comparison with the currently-known fastest pulsars
PSR J1748-2446ad in the globular cluster Terzan 5
with spin frequency $F=716\,$Hz
and the X-ray transient XTE J1739-285 with $F=1122\,$Hz,
which are still far below the theoretical limit \cite{Hessels06}.

Comparing Keplerian and static sequences,
rotations increase the maximum mass
and equatorial radius substantially.
The maximum mass increases by about 20\%
from the static to the Keplerian sequence
\cite{Cook94,Lasota96,Breu16,Khadkikar21},
and the radius by a similar amount.
Note that the full dots on the Keplerian sequences in the figure
indicate the maximum gravitational mass points,
which may slightly deviate from the onset of instability \cite{Wei17},
which is for the static sequence determined by the condition
$\partial M/\partial \rho_c=0$,
i.e., at the maximum mass.
This criterion has to be generalized
by the secular axisymmetric instability
in the rotating case \cite{Haensel07}.

\begin{figure*}[t]
\vskip-0mm\hskip0mm
\centerline{\includegraphics[width=1.0\textwidth]{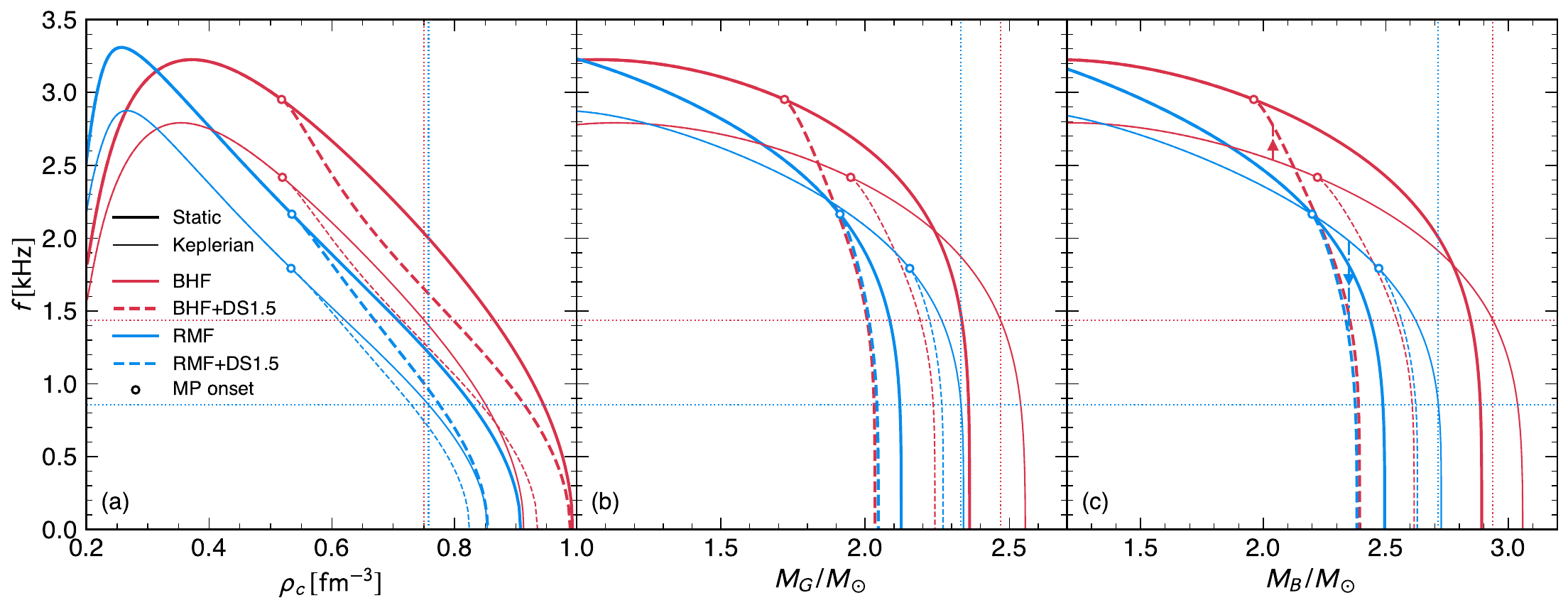}}
\vskip-5mm
\caption{
The fundamental (Q)RO frequency $f = \om/2\pi$
as a function of
central baryon number density $\rho_c$ (a),
gravitational mass $M_G$ (b), and
baryonic mass $M_B$ (c)
with various EOSs.
Thick (thin) curves denote static (Keplerian) sequences.
Open dots indicate the onset of the MP.
Dotted lines indicate borders on the validity
of the perturbative approach
(in the top left delimited quadrants).
The vertical lines in (c) show example spin-down trajectories.
\hj{}
}
\label{f:f-MG}
\end{figure*}

\begin{figure}[t]
\vskip-1mm\hskip0mm
\centerline{\includegraphics[width=0.5\textwidth]{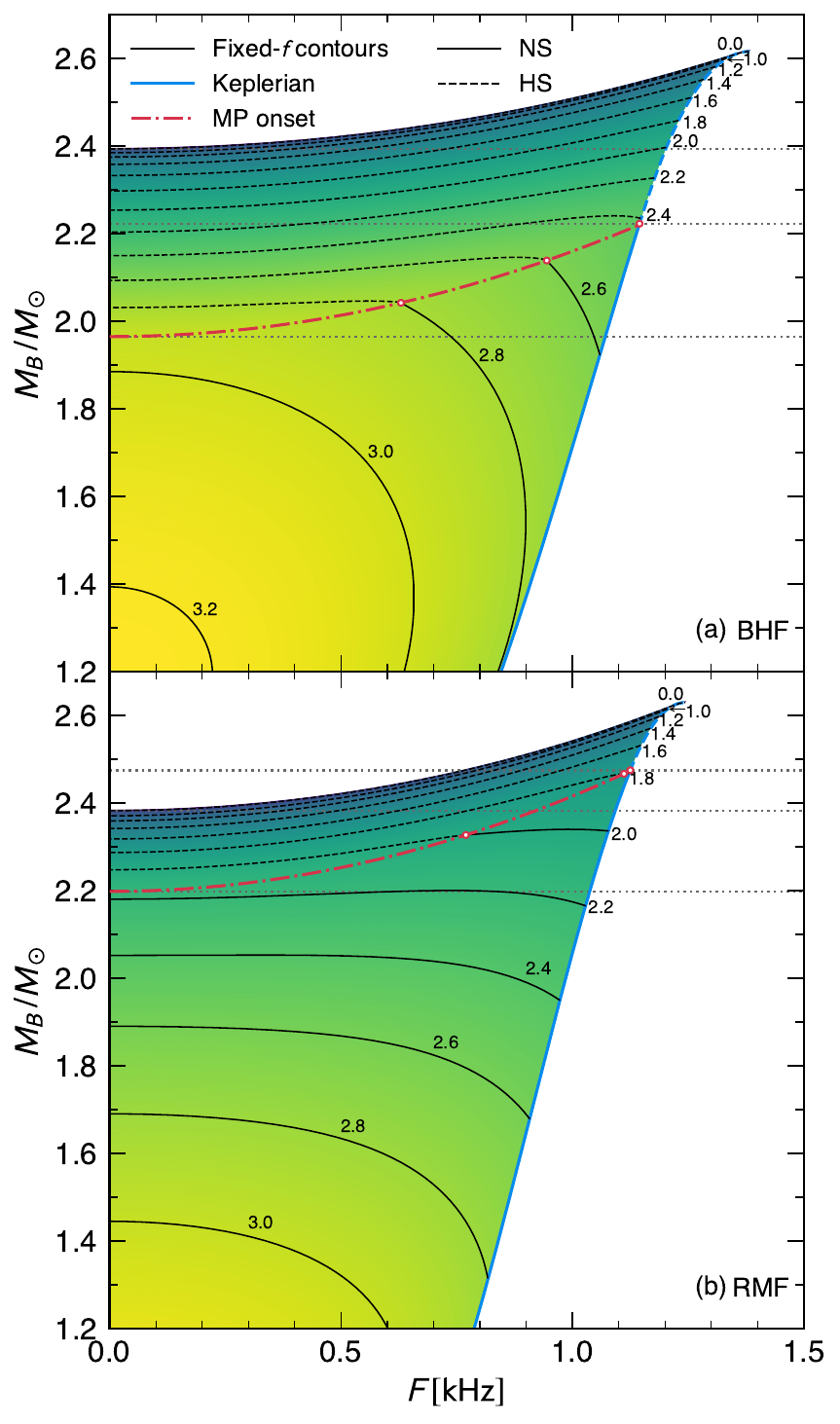}}
\vskip-5mm
\caption{
The $f=\om/2\pi$ contours
($f$ in kHz indicated near the thin black curves)
in the $[M_B,F=\Om/2\pi]$ plane
for BHF (a) and RMF (b) EOSs,
up to the Keplerian frequency (thick blue curve).
The dash-dotted red curve indicates the HQ transition.
Shown are only the HS frequencies beyond that transition.
The solid (dashed) curve segments represent the pure NS (HS) branches.
The horizontal dotted lines indicate some limiting evolution trajectories
discussed in Secs.~\ref{s:res}C,D.
\hj{}
}
\label{f:F-MB}
\end{figure}

\subsection{Stellar oscillations}

Fig.~\ref{f:f-MG}(a) shows the
$\ord(\Om^2)$ fundamental (Q)RO frequency $f = \om/2\pi$,
Eq.~(\ref{e:sig}),
along both static and Keplerian sequences,
as a function of the central baryon number density $\rho_c$.
In general,
as the star's equatorial radius increases due to rotation,
its matter becomes more dilute,
leading to a decrease of the QRO
frequency as the rotation rate
approaches the Keplerian limit.
The RMF EOS with the smaller $\Gamma$
yields lower frequencies than the BHF EOS.

As explained above,
for a given $\rho_c$,
the difference between the static and Keplerian curves
is due to the rotational correction $\om_2^2$,
Eq.~(\ref{e:sigma2}),
to the static QRO frequency $\om_0^2$,
Eq.~(\ref{e:ddeltapdr}).
With increasing $\rho_c$,
$\om_2^2$ becomes of same size as $\om_0^2$
and perturbation theory ceases to be valid.
Therefore,
to empirically estimate the validity of perturbation theory,
the dotted lines in the figure indicate where the static frequency $\om_0^2$
is changed by 50\%.
This occurs well beyond the MP onset configurations.
With rotation frequency $\Om$ decreasing from the Keplerian frequency,
the dotted lines shift to the bottom/right,
extending the regime where perturbation theory applies.

Fig.~\ref{f:f-MG}(b) shows the same curves
as a function of the gravitational mass $M_G$.
ROs are commonly used as a criterion for determining stellar stability,
where in our model the stability condition
$\partial M / \partial \rho_c \geq 0$
indeed corresponds to the onset of unstable oscillations
for static stars at $\Om = 0$ \cite{Sun21,Sun25}.
However, for Keplerian sequences,
the QRO frequencies derived from the slow-rotation approximation
do not correctly recover the critical point of stability, i.e.,
$f=0$ in Fig.~\ref{f:f-MG}(b)
does not exactly coincide with $\mmax$ on the Keplerian sequence
in Fig.~\ref{f:m-r}(b),
but the difference is small.
The nondegenerate perturbation theory,
which underlies expressions such as Eq.~(\ref{e:sigma2}),
is inapplicable in this case.
Its domain of validity is restricted to regimes
where the unperturbed oscillation frequency $\om_0$
is large compared to the rotational frequency $\Om$,
i.e., $\om_0^2 \gg \Om^2$ \cite{Hartle75}.
This implies that the results obtained from perturbation theory
are reasonably reliable when $\om_0$ is relatively large,
i.e., when the gravitational mass is not close to $\mmax$.

To visualize the evolution of the QRO frequency during stellar spin-down,
we show in Fig.~\ref{f:f-MG}(c)
that frequency as a function of the baryonic mass $M_B$,
which is approximately conserved during such spin-down.
In this work,
we investigate in particular the evolution of the QRO frequency
during the spin-down of a rapidly-rotating pure NS
into a slow-rotating hybrid star,
as for example represented by the vertical lines at
$M_B=2.05\ms$ (BHF) and
$M_B=2.35\ms$ (RMF).

In order to investigate this process in detail,
we show in Fig.~\ref{f:F-MB}
the $f = \om/2\pi$ contours
in the $[M_B, F=\Om/2\pi]$ plane
for the BHF (a) and RMF (b) EOSs,
up to the Keplerian frequency.
The dash-dotted red curve denotes the MP onset.
Since $M_B$ is approximately conserved during spin-down,
each evolutionary sequence corresponds
to a horizontal track from large to small $F$ in this plane.
For the BHF (RMF) EOS,
the MP-onset curve extends
from $M_B \approx 1.96(2.20)\ms$ at $F=0$
to $M_B \approx 2.22(2.47)\ms$ at $F \approx 1.15(1.13)\khz$.
Therefore,
three mass regimes can be directly identified:
\\--
For $M_B \lesssim 1.96(2.20)\ms$,
the fixed-$M_B$ tracks remain below the MP-onset curve,
and the star evolves entirely along the pure NS branch.
\\--
For $1.97(2.20)\ms \lesssim M_B \lesssim 2.22(2.47)\ms$,
these sequences contain a spin-down-induced phase transition
and can evolve to stable nonrotating HSs for the BHF EOS,
but only up to $M_B \lesssim 2.38\ms$ for the RMF EOS,
beyond which there will be a transition to a BH.
\\--
For $M_B \gtrsim 2.22(2.47)\ms$,
up to the maximum HS masses $M_B \approx 2.62(2.63)\ms$,
the rapidly-rotating configurations are already HSs.
In this regime,
the BHF sequences can evolve to massive static HSs
only for $2.22\ms \lesssim M_B \lesssim 2.40\ms$.
Above $M_B \approx 2.40(2.47)\ms$ for the BHF (RMF) EOS,
no stable nonrotating counterpart exists
at the same baryonic mass,
implying collapse to a BH.
\\
The detailed evolution of $f(F)$ along these different classes
of fixed-$M_B$ tracks is discussed in the next section.

\subsection{Evolution}

\begin{figure*}[t]
\vskip-5mm
\centerline{\includegraphics[width=1.0\textwidth]{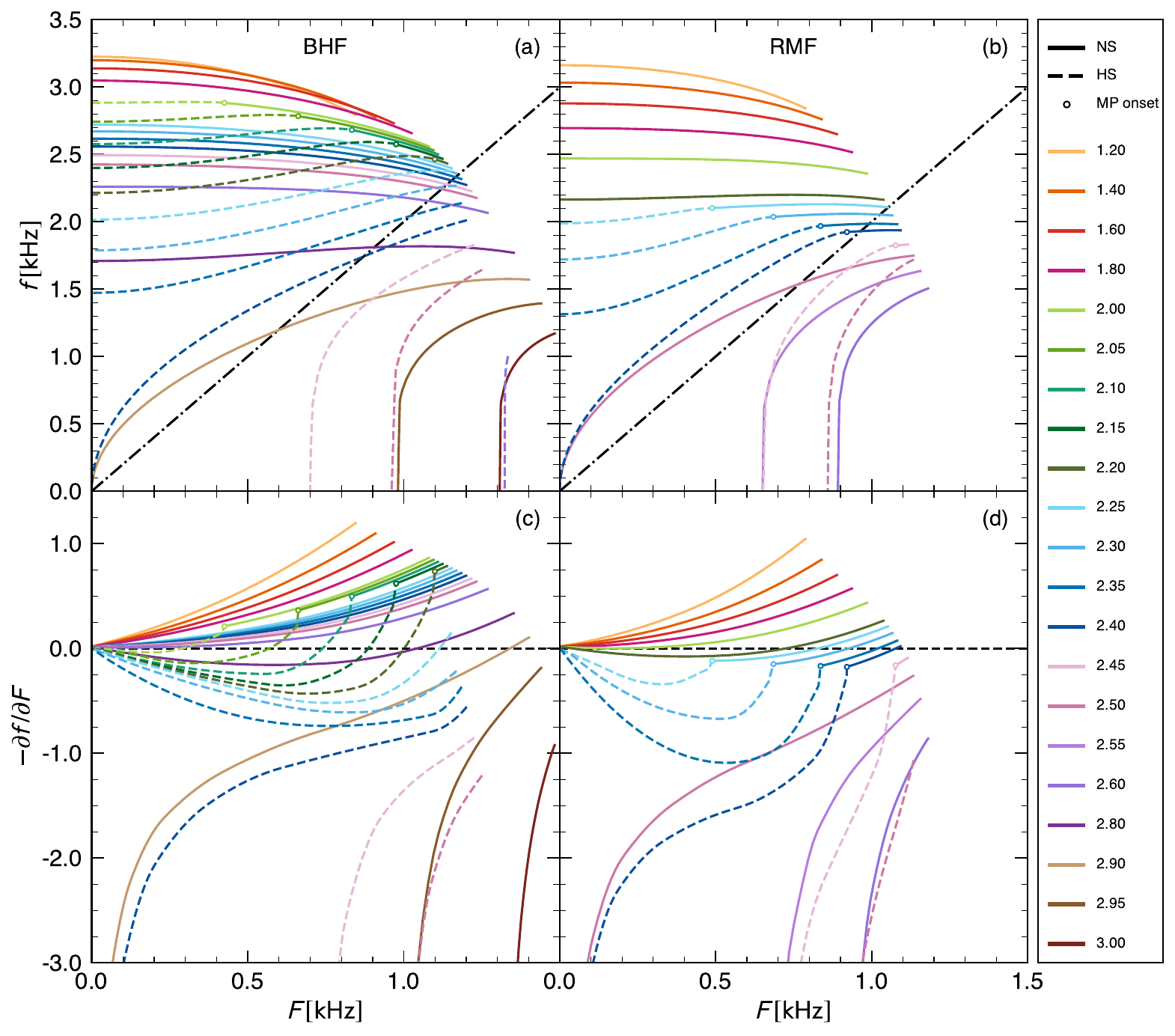}}
\vskip-4mm
\caption{
Fundamental QRO frequency $f=\omega/2\pi$ (a,b)
and its derivative with respect to rotation frequency $F=\Om/2\pi$ (c,d)
vs $F$,
for several fixed baryonic masses $M_B/\ms$,
obtained with the BHF (a,c) and RMF (b,d) EOSs.
The solid (dashed) curves correspond to NSs (HSs),
and open dots indicate the bifurcation points of pure NSs and HSs.
Diagonal lines indicate the $f=2F$ limit.
}
\label{f:f-F}
\end{figure*}

\begin{figure*}[t]
\vskip-5mm
\centerline{\includegraphics[width=1.0\textwidth]{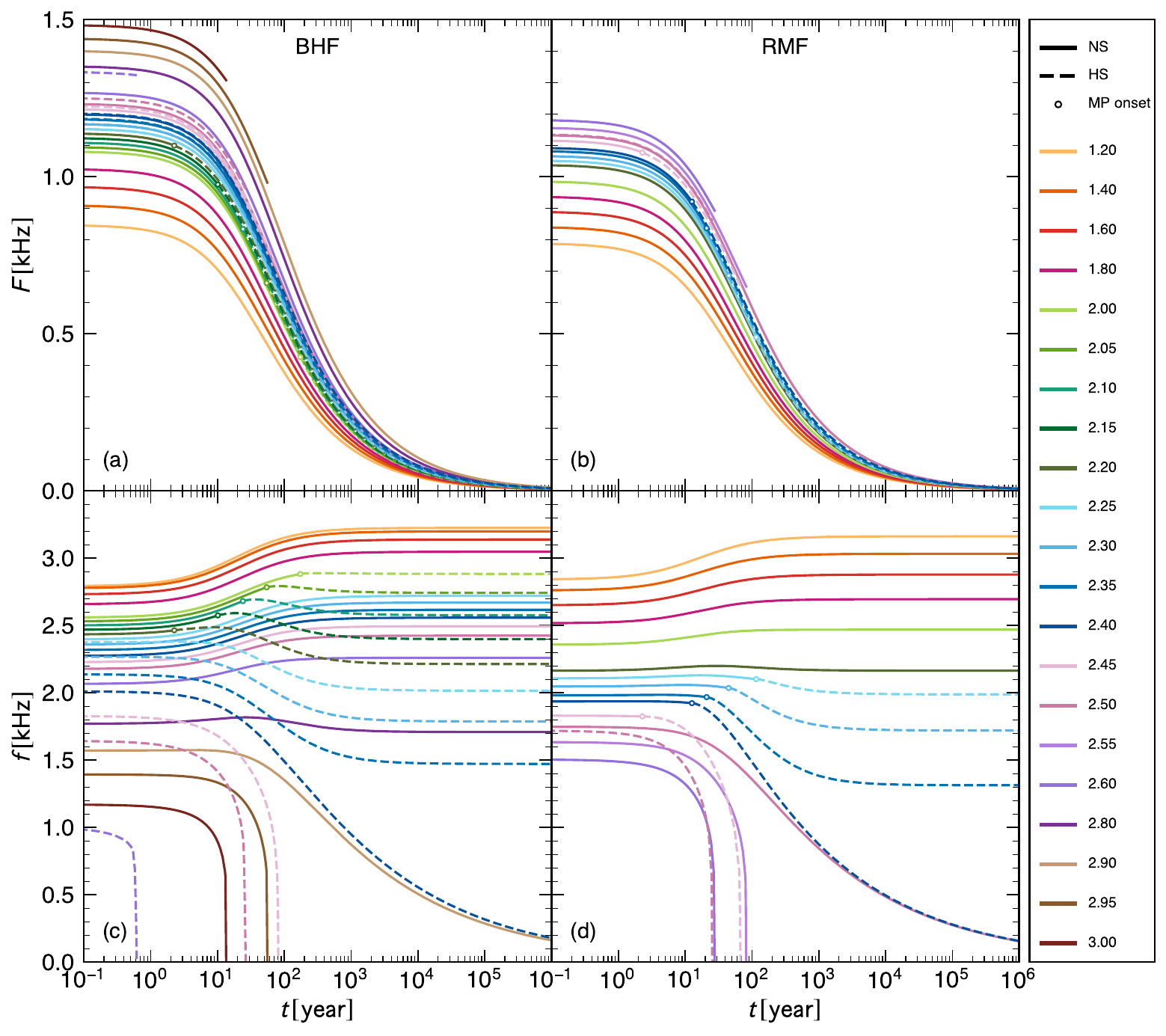}}
\vskip-4mm
\caption{
The time evolution of rotation frequency $F$ (a,b)
and fundamental QRO frequency $f$ (c,d) during spin-down
for several fixed values of baryonic mass $M_B/\ms$
as in Fig.~\ref{f:f-F},
obtained with the BHF(a,c) and RMF (b,d) EOSs.
The solid (dashed) curves correspond to NSs (HSs),
and open dots indicate the bifurcation points of pure NSs and HSs.
}
\label{f:f-t}
\end{figure*}

Fig.~\ref{f:f-F}(a,b) shows
the fundamental (Q)RO frequency $f=\om/2\pi$
as a function of rotation frequency $F=\Om/2\pi$
for several fixed values of baryonic mass $M_B$,
as derived from Fig.~\ref{f:F-MB}.
As just outlined,
there are several classes of evolution $f(F)$,
depending on $M_B$:
\\--
For low $M_B$
[$M_B \lesssim 1.96(2.20)\ms$ for BHF(RMF)]
there are no HS configurations
and $f$ increases steadily with decreasing $F$,
to values of more than 3~kHz for light stars.
\\--
More massive static HSs
[$M_B \approx 1.96\text{-}2.22(2.20\text{-}2.47)\ms$ for BHF(RMF)]
may instead evolve from initial fast-rotating pure NSs
and experience a MP onset
(open markers in Figs.~\ref{f:F-MB} and \ref{f:f-F})
during spin-down.
At this point, the initial increase of $f$ with decreasing $F$ in the NS
turns into a pronounced decrease in the HS.
This is because the static maximum mass of the HS
(where $f=0$)
is smaller than that of the pure NS.
\\--
The most massive static HSs in the BHF model
[$M_B \approx 2.22\text{-}2.40\ms$]
are not accessible from rapidly-rotating pure NSs,
but must instead evolve from HSs.
In this case $f$ decreases steadily with decreasing $F$,
eventually dropping to very low values for the static star
(outside the validity of the perturbative approach).
\\--
For even larger
$M_B \approx 2.40\text{-}2.62(2.38\text{-}2.63)\ms$,
there are no stable static HS configurations,
and an initially fast-rotating HS will transit to a BH during spin down
(where $f=0$).
The most massive HSs with the BHF(RMF) models are
$M_B=2.62(2.63)\ms$ with $F_K=1.39(1.25)\khz$,
highly unstable.

If instead HSs do not exist,
there are fast-rotating NSs up to $M_B=3.06(2.73)\ms$,
which exhibit similar phenomena as heavy HSs during their spin-down,
namely a strong decrease of the oscillation frequency
towards $f=0$ of the static star,
and a collapse to a BH is possible also here.
As mentioned in Sec.~\ref{s:res}C,
we note that such slowly oscillating but rapidly rotating stars
are not reliably modeled in the perturbative approach,
and the results should be understood as qualitative.
A rough guidance is provided by the $f=2F$ diagonal lines in the figure.

In summary, a decrease of $f$ during spin-down identifies the star
as being a relatively massive HS,
with a larger decrease
indicating a configuration closer to the static maximum mass.
However, the decline in $f$
from HSs may be indistinguishable from that of more massive pure NSs,
posing a challenge to discriminate both cases.
For this purpose,
the presence of a phase transition during this evolution
leads to a pronounced change of slope of the $f(F)$ curve,
which might be observable,
also due to the associated relatively short timescales.

To clearly illustrate the bifurcation point between pure NSs and HSs,
we show in Fig.~\ref{f:f-F}(c,d)
the derivative $-\partial f / \partial F$
as a function of rotation frequency.
It can be seen that
for pure NSs this quantity decreases with increasing baryonic mass.
It is initially positive for low masses
and becomes negative as the maximum mass is approached.
If the HQ phase transition occurs during the evolution,
there is an associated distinct change of $\partial f / \partial F$,
much more apparent than in panels (a,c).
Consequently,
a future detection of a kink point of $\partial f / \partial F$
during spin-down,
followed by a significant increase,
could signal the occurrence of a HQ phase transition inside the star.

In order to assess the timescale over which a HQ phase transition
could be observed during the long-term spin-down of a NS,
we employ a simple spin-down model to compute the time evolution of
the frequencies $F$ and $f$.
Assuming that the spin-down is purely due to electromagnetic dipole radiation,
the change of rotation frequency $\Om$ with time $t$ is
\cite{Shapiro83}
\be
 \frac{d\Om}{dt} = - \frac{B_p^2 R(\Om)^6 \Om^3}{6I(\Om)} \:,
\label{e:dOmdt}
\ee
where $B_p$ is the dipole field strength at the poles,
$R(\Om)=R(0)+\delta r_0(R)$ is the mean radius, and
$I(\Om)=I(0)+\delta I$ is the moment of inertia,
both at $\ord(\Om^2)$.
A contribution due to $dI/d\Om$ in Eq.~(\ref{e:dOmdt}) is neglected.
Since the magnetic field strength of a typical pulsar
is of the order of $10^{11}\text{-}10^{13}\,$G
\cite{Enoto19},
we adopt a representative value of $B_p=10^{12}\,$G
and assume that the field remains constant throughout the spin-down process.

Fig.~\ref{f:f-t}(a,b) shows the time evolution of the rotation frequency $F$,
where the Keplerian frequency from Eq.~(\ref{e:f_K})
is taken as the initial rotation frequency.
Compared to NSs with the same baryonic mass,
HSs have a slightly smaller $dF/dt$
due to their slightly smaller radius.
After approximately $10^6$ years,
the rotational frequency of stable stars
has already decreased to below $0.01\khz$.

Fig.~\ref{f:f-t}(c,d) shows the corresponding time evolution
of the QRO frequency $f$ during spin-down.
In this figure,
the evolution of $f$ for pure NSs can be seen more clearly.
For low-mass stars,
$f$ increases monotonically with time.
As the baryonic mass increases,
this behavior transitions to a slight initial increase
followed by a decrease.
For stars approaching the maximum mass,
the evolution becomes a monotonic decrease.
The phase transition to a HS occurs earlier for larger baryonic masses.
In the figure,
the earliest transition,
for BHF at $M_B=2.2\ms$,
takes place approximately two years after the star's birth,
while for $M_B=2.0\ms$ at about 200 years.
But any time is possible, depending on small changes of $M_B$.
Note also that the figure assumes formation of the star
at Keplerian frequency,
which is not likely to occur.
A phase transition occurring
in fast-rotating young pulsars
implies richer observable signatures.
In particular,
the kink behavior associated with the HQ transition
shown in Fig.~\ref{f:f-F}(c,d) could be detected
over an observational timescale of decades to centuries.
This is much longer than the local microscopic conversion timescales
discussed in Ref.~\cite{Lugones16},
where the process is described as proceeding in two steps,
with deconfinement first occurring
on a strong-interaction timescale of $\sim10^{-24}\,{\rm s}$,
followed by weak-interaction equilibration
on a timescale of $\sim10^{-8}\,{\rm s}$.

\section{Conclusions}
\label{s:end}

In this work,
we investigated the QROs of rotating NSs
within the slow-rotation perturbative approximation,
considering both pure NSs and HSs.
The EOS is constructed using the BHF approach or RMF theory for NM,
together with the DSM for QM,
assuming a phase transition governed by the Gibbs construction.
For a fixed central density,
the rotation effect reduces the fundamental QRO frequency
for both pure NSs and HSs.
Considering the validity of the slow-rotation approximation,
we refrain from using the QRO to discuss stellar stability.

We also investigated the phase transition induced by the spin-down of pulsars
with a constant baryonic mass.
With the star slowing down after birth,
its central density rises.
Once it crosses the critical density for phase transition,
QM appears at the center of the star.
As the star continues to spin down,
the QM region expands over time.
We presented the fundamental QRO frequency $f$
as a function of rotation frequency $F$
for several fixed baryonic masses,
where the appearance of QM in more massive configurations
leads to a pronounced decrease of $f$ during spin-down.
However, a similar behavior is also observed
in pure NSs approaching the static maximum mass,
making the evolutionary behavior of $f$
indistinguishable between HSs and massive pure HSs.
This degeneracy arises because $f=0$
at the static maximum-mass configurations,
so any star approaching this limit experiences a greater decrease in
$f$ during spin-down.
Additionally,
QM softens the EOS,
resulting in a lower maximum mass for static HSs.
Hence, HSs exhibit the pronounced decrease
at lower masses compared to pure NSs.

To clearly illustrate the bifurcation point
between pure NSs and HSs during spin-down,
we showed the derivative of $f$ with respect to the rotation frequency,
$\partial f / \partial F$,
as a function of rotation frequency.
At the onset of QM during spin-down,
this quantity exhibits a kink followed by a steep rise,
and provides thus a distinct observational signature
that differentiates HSs from their pure NS counterparts.
Translated into the time evolution of the fundamental QRO frequency
within a simplified spin-down model,
this kink behavior of $\partial f / \partial F$,
associated with the HQ transition,
could be detectable over an observational window of decades to centuries.

To investigate QROs in rotating NSs,
the features of more realistic environments,
e.g., temperature and magnetic field,
should also be included.
Furthermore,
to ensure the accuracy of calculations under rapid rotation,
a full general relativistic framework is also necessary.
We leave these studies to future work,
together with those employing other EOSs.

\begin{acknowledgments}

Zi-Yue Zheng and Xiao-Ping Zheng are supported by
the National Natural Science Foundation of China (Grant No.~12033001)
and
the National SKA Program of China (Grant No.~2020SKA0120300).
Jin-Biao Wei, Huan Chen, and Ting-Ting Sun acknowledge financial support
from the National Natural Science Foundation of China (Grant No.~12205260).

\end{acknowledgments}

\newcommand{\epja}{Euro. Phys. J. A}
\newcommand{\aap}{Astron. Astrophys.}
\newcommand{\apjl}{Astrophys. J. Lett.}
\def\jcap{Journal of Cosmology and Astroparticle Physics}
\def\jcap{JCAP}
\newcommand{\mnras}{Mon. Not. R. Astron. Soc.}
\newcommand{\nphysa}{Nucl. Phys. A}
\newcommand{\physrep}{Phys. Rep.}
\newcommand{\plb}{Phys. Lett. B}
\newcommand{\ppnp}{Prog. Part. Nucl. Phys.}
\bibliography{nsqro}

\end{document}